\documentclass[manuscript]{acmart}

\AtBeginDocument{%
  \providecommand\BibTeX{{%
    \normalfont B\kern-0.5em{\scshape i\kern-0.25em b}\kern-0.8em\TeX
    
    }}}
    
\usepackage{ulem}

\newcommand{\ra}[1]{\renewcommand{\arraystretch}{#1}}

\setcopyright{acmcopyright}
\copyrightyear{2025}
\acmYear{2025}
\acmConference[GROUP '25]{GROUP '25: ACM International Conference on Supporting Group Work}{January 12-15, 2025}{Hilton Head, USA}

\usepackage{multirow}
\begin{document}

\title[Building Solidarity Amid Hostility]{Building Solidarity Amid Hostility: Experiences of Fat People in Online Communities}

\author{Blakeley H. Payne*}
\email{blakeley.hoffman@colorado.edu}

\affiliation{
   \institution{University of Colorado Boulder}
\city{Boulder}
\state{CO}
\country{USA}
}

 \author{Jordan Taylor*}
\email{jordant@andrew.cmu.edu}
 \affiliation{
   \institution{Carnegie Mellon University}
  \city{Pittsburgh}
  \state{PA}
 \country{USA}
 }

 \author{Katta Spiel}
 \email{katta.spiel@tuwien.ac.at}
 \affiliation{
  \institution{TU Wien}
\city{Vienna}
\country{Austria}
}

\author{Casey Fiesler}
\email{casey.fiesler@colorado.edu}
\affiliation{
   \institution{University of Colorado Boulder}
\city{Boulder}
 \state{CO}
\country{USA}
}

\begin{abstract}
    
Online communities are important spaces for members of marginalized groups to organize and support one another. To better understand the experiences of fat people — a group whose marginalization often goes unrecognized — in online communities, we conducted 12 semi-structured interviews with fat people. Our participants leveraged online communities to engage in consciousness raising around fat identity, learning to locate ``the problem of being fat'' not within themselves or their own bodies but rather in the oppressive design of the society around them. Participants were then able to use these communities to mitigate everyday experiences of anti-fatness, such as navigating hostile healthcare systems. However, to access these benefits, our participants had to navigate myriad sociotechnical harms, ranging from harassment to discriminatory algorithms. In light of these findings, we suggest that researchers and designers of online communities support selective fat visibility, consider fat people in the design of content moderation systems, and investigate algorithmic discrimination toward fat people. More broadly, we call on researchers and designers to contend with the social and material realities of fat experience, as opposed to the prevailing paradigm of treating fat people as problems to be solved in-and-of-themselves. This requires recognizing fat people as a marginalized social group and actively confronting anti-fatness as it is embedded in the design of technology.

\end{abstract}

\begin{CCSXML}
<ccs2012>
   <concept>
       <concept_id>10003120.10003130.10011762</concept_id>
       <concept_desc>Human-centered computing~Empirical studies in collaborative and social computing</concept_desc>
       <concept_significance>500</concept_significance>
       </concept>
   <concept>
       <concept_id>10003456.10010927</concept_id>
       <concept_desc>Social and professional topics~User characteristics</concept_desc>
       <concept_significance>500</concept_significance>
       </concept>
 </ccs2012>
\end{CCSXML}

\ccsdesc[500]{Human-centered computing~Empirical studies in collaborative and social computing}
\ccsdesc[500]{Social and professional topics~User characteristics}

\keywords{Fat People, Online Communities, Consciousness Raising}

\maketitle

\section{Introduction}

In March 2020, online news organization The Intercept published leaked content moderation policy documents from TikTok, revealing guidelines the company had put forth to censor ``undesirable'' users on the platform. Among the list of undesirable users were those TikTok described as ``chubby,'' ``obese,'' having an ``obvious beer belly,'' or an ``abnormal body shape'' \cite{invisible_censorship}. Later that year, in August, reporting from a variety of platforms documented similar content moderation problems on Instagram. The hashtag \#IWantToSeeNyome brought attention to the way photographs of Black, fat women on Instagram - and photographs of the model Nyome Nicholas Williams in particular - were often flagged as violating Instagram's nudity policies \cite{instagram_fat_discrimination}. Alongside this greater public recognition, HCI scholars are beginning to study the ways in which anti-fatness manifests in Reddit \cite{rfatpeoplehate}, online anti-vax communities \cite{koltai}, and social media diet advertisements \cite{gak2022distressing}. In addition to studying \textit{anti-fatness}, HCI scholars are increasingly calling for more research centering the lived experiences of fat people \cite{payne2023ethically, jonas2024we, sobey2024conceptualising}. Nevertheless, it is still rare for fat people to be recognized as a marginalized social group, and research on the experiences of fat people in HCI remains scarce \cite{sobey2024conceptualising}.

Despite these challenges associated with anti-fatness, fat people have a long history of appropriating technology \cite{eglash2004appropriating} — even those not necessarily designed with fat people in mind — to form community. In the early 2000s, the ``Fat-O-Sphere'' formed from a series of interconnected blogs written by and about fat people. These blogs challenged dominant narratives of fat people and acted as spaces for fat people to make friends with one another \cite{Harding_Kirby_2009a}. Since then, the ``fatosphere'' has expanded to include new social media platforms and the online communities they facilitate.

In the tradition of prior research on how marginalized groups negotiate the harms and benefits of online communities \cite{pinter2021entering, klassen2021more}, in this exploratory study we attend to the specific experiences of fat people in online communities. We conducted twelve semi-structured interviews with fat people and explored three overarching questions:

\begin{enumerate}
    \item How do fat people benefit from participating in online communities?
    \item What harms do fat people experience in online communities?
    \item What strategies do fat people employ to negotiate these harms and benefits?
\end{enumerate}

Below, we report how online communities provided space for our participants to re-conceptualize their fat identity, helping them better understand the machinations of fat oppression and navigate daily life. However, we found that online communities are often particularly hostile toward fat people, leading our participants to contend with both hostile users and hostile design. In response, participants deployed a variety of strategies, such as satirizing trolls, individually checking-in on targets of harassment, and building one's own community. In light of our findings, we encourage HCI researchers to attend to the politics of how we know fat people. We also call on researchers to design against anti-fatness, which requires contending with the material and social aspects of fat embodiment and identity. Finally, we provide recommendations for researchers seeking to design better online communities for fat people, such as improving content moderation, supporting selective visibility, and designing for fat community building.

\section{Background on Fat People, Fat Oppression, and Fat Liberation}

    We intentionally use the word ``fat'' instead of medicalizing language such as ``obesity'' or ``overweight.'' We do this to align ourselves with fat activists and fat studies scholars who have reclaimed the word both as a neutral bodily descriptor (e.g., similar to descriptors such as ``tall'' or ``has brown hair'') and facet of identity \cite{wann2009fat}. This reclamation parallels the way other marginalized communities have reclaimed language leveraged against them historically, such as the word ``queer'' among LGBTQIA+ individuals and ``crip'' within the disability justice movement. We discuss our choice to use the word ``fat'' in greater detail in our prior work on how to ethically engage fat people in HCI research \cite{payne2023ethically}. 

    Fat people experience substantial social marginalization. Around the world, discriminatory policies prevent fat people from immigrating \cite{pause2021frozen} to other countries or accessing particular healthcare (such as fertility treatments, orthopedic surgeries, or organ transplants) \cite{pause2014another}. Hostile architecture, such as small seating or seats with low weight capacity, can prevent fat people from accessing public spaces such as restaurants, airplanes, or university classrooms \cite{wann2009fat}. Fat people are often mistreated by doctors \cite{no-health-no-care, paine2021fat, phelan2015impact} as well as negatively stereotyped as lazy, unintelligent, and unattractive \cite{crandall1994prejudice, farrell2011fat} and poorly depicted in media \cite{kyrola2016weight}. One way the media dehumanizes fat people is via the ``headless fatty'' phenomena which refers to how fat people are often filmed without their consent from the neck down (i.e. ``headless'') \cite{cooper2007headless}. 
    
    Knowledge production is another arena where fat oppression is reproduced \cite{cooper2014no}. Within the field of fat studies, scholars prioritize fat epistemology, or ways of knowing which center the expertise of fat people. This is in response to fat people historically being excluded from knowledge production arenas and processes, such as in academia and academic research \cite{pause2021fattening, pause2020ray, cooper2020elevating}. As fat activist and scholar Rachel Fox points out, the dominance of ``obesity epidemic'' discourses has narrowed both what is known — and considered important to know — about fat people \cite{value-fat}. 

    Beginning in the early 1970s, fat people — primarily white, queer women involved in the U.S. women's liberation movement— began to organize around these issues. Groups such as the Fat Underground, the Radical Feminist Therapy Collective, the National Association to Advance Fat Acceptance gathered, wrote, and marched to draw attention to the ways in which fat oppression influences all spheres of life \cite{simic2016fat, friedman2019thickening, schoenfielder1983shadow}. In 1973, members of the Fat Underground published \textit{The Fat Liberation Manifesto}, outlining how institutions like the diet industry, academia, and medicine uphold fat oppression \cite{fat-lib-manifesto}. Here, \textit{fat liberation} refers to a future where fat people are granted full and equal access to all parts of society. More specifically, fat liberation rejects the notion that fat people must diet or shrink their bodies to access resources or spaces such as public transportation, employment, or healthcare, and that fat people deserve to be treated with dignity and respect \cite{cooper2009maybe}. At the same time, like any political movement, people who would describe themselves as fat liberationists or fat activists can hold many competing and overlapping beliefs. While often thought to be a uniquely American pursuit, fat liberation efforts and organizing exist worldwide \cite{cooper2009maybe, casado2020m, mibellibelle, puhakka2021sofie, hynna2019feel}. 

    Fat liberation is distinct from several other framings of fatness. In particular, fat liberation is often conflated with body positivity. Where body positivity focuses on a person's internal experiences and feelings of their body, fat liberation focuses on tearing down external structures that prevent fat people from accessing society \cite{cooper2010fat, gibson2022health}. For example, while a body positive outlook might encourage a fat person to feel confident while wearing a particular outfit, fat liberation instead focuses on the extremely limited access fat people have to clothing at all, and encourages retailers to sell clothing in shapes and sizes that fit fat people. Fat liberation is also distinct from the Health at Every Size framework, a set of principles intended to support clinicians in providing better care for fat patients \cite{asdah-principles, gibson2022health}. In contrast, fat liberation emphasizes that the right to be treated justly is not contingent on performing health behaviors. We approach our work from a fat liberationist perspective informing, for instance, our decision to use the word ``fat'' in this paper.

\section{Related Work}

\subsection{Consciousness Raising in Online Communities}

    Much of early fat activist efforts centered around the process of feminist consciousness raising \cite{schoenfielder1983shadow, farrell2019origin}. Consciousness raising activities saw women gather in local groups to discuss their individual lived experiences as a means to identify common patterns of mistreatment, develop theories of gender-based oppression, and engender collective political action \cite{fraser2014rethinking, fricker2007epistemic, sarachild2000program}. It was within these consciousness raising sessions that the language to describe post-partum depression and sexual assault was invented \cite{fricker2007epistemic}. For US fat activists in the 1970s, consciousness raising discussions often elucidated the harms of dieting, size-based employment discrimination, mistreatment and neglect by medical professionals, or the inaccessibility of public spaces like movie theaters or airplanes due to undersized seating. Through these discussions, fat activists began to reject the notion that they individually needed to change their body in order to merit equitable access to spheres of society such as employment, healthcare, or leisure - rather, society needed to change \cite{schoenfielder1983shadow}. Within HCI, D'Ignazio et al. discussed how feminist hackathons can function as important spaces for consciousness raising \cite{d2020personal}. 
    
    While not always described as ``feminist consciousness raising,'' HCI scholarship has long been interested in the ways online communities facilitate the construction of knowledge. For instance, research drawing on sensemaking \cite{russell1993cost} — the ongoing process of trying to construct a working understanding of the world — has explored how people collectively understand their health \cite{mamykina2015collective, young2019girl} and audit algorithmic systems \cite{shen2021everyday}. Prior online community research has also emphasized the highly political nature of collective knowledge construction, such as Twitter activism countering police narratives \cite{jackson2015hijacking} and raising awareness of sexual violence \cite{jackson2020hashtagactivism, moitra2020understanding}. This form of knowledge contestation often takes place through storytelling. Dimond et al. discussed how victims of harassment benefited from sharing their stories online, allowing them to ``shift'' from seeing their experience as an individual one to part of a collective \cite{dimond2013hollaback}. As modeled by Ogbonnaya-Ogburu et al. in their work on critical race theory, storytelling can be used as ``a means to explore oppression'' \cite{ogbonnaya2020critical}.

    Online communities have long been spaces for marginalized groups to build counter-knowledge and push back against dominant narratives or ways of knowing. For instance, researchers have explored how people use online communities to decolonize their identities \cite{dosono2020decolonizing, das2022collaborative}, such as discussions in Bengali Quora challenging national categories imposed by British colonizers \cite{das2022collaborative}. Similarly, queer online communities are also more than mere places where people learn about queer identity. Rather, members of these communities actively construct their own understandings of queerness, often pushing back against dominant cultural narratives \cite{dym2019coming, simpson2021you}. At the same time, online identity construction is often mediated by algorithmic systems that may elevate the most normative identity presentations \cite{simpson2021you, karizat2021algorithmic}. Nevertheless, marginalized people can play an active role in countering algorithmically elevated normative identity representations via folk theorizing \cite{devito2018people, karizat2021algorithmic}.

\subsection{The Design of Online Communities \& Marginalized Groups}

    Online communities are often designed in ways that fail to meet the needs of marginalized groups \cite{simpson2022tame}. For instance, real name policies may not account for those whose names do not match those on official government documents or may wish to use different names with different audiences \cite{haimson2016constructing}. Marginalized groups often experience identity-based harassment in online communities \cite{mandryk2023combating, gray2014race, heung2024vulnerable, schulenberg2023creepy}. These experiences can silence the voices of marginalized groups and prevent people from participating in, for instance, gaming \cite{gray2014race} or VR communities \cite{schulenberg2023creepy}. Moreover, marginalized groups can also be harmed by punitive content moderation systems \cite{haimson2021disproportionate, salehi2023sustained}. Social media platforms often fail to remove hate speech or target harassment toward marginalized groups \cite{salehi2023sustained}, while at the same time subjecting members of marginalized groups to disproportionate content removal \cite{haimson2021disproportionate}. To better support marginalized groups, HCI scholars have begun to explore alternatives to punitive content moderation systems, such as restorative justice approaches \cite{salehi2023sustained, xiao2023addressing} and intersectional approaches built on care and power \cite{gilbert2023towards}.

    Members of marginalized groups typically navigate online spaces that facilitate representational harms. Often discussed within the machine learning and AI fairness literature, representational harms occur when ``when some social groups are cast in a less favorable light than others, affecting the understandings, beliefs, and attitudes that people hold about these social groups'' \cite{wang2022measuring, barocas2017problem}.  Infamous examples of representational harms include the harm done when Google Photos mistakenly classified images of Black people as ``apes,'' or when Google Search produced hypersexualized results for queries related to girls of color \cite{noble2018algorithms}. Shelby et al. identify five different ways representational harms can occur, including  the stereotyping, demeaning, erasing, and alienating of social groups, as well as denying the opportunity to self-identify, or by reifying essentialist categories \cite{shelby-reprsentational}. A growing number of works have examined how representational harms are facilitated within the design of online communities. For example, programmatic social media advertising \cite{sampson2023representation} and algorithmic social media feeds \cite{simpson2021you} can promulgate stereotypical representations of LGBTQ+ people. Research on Black \cite{harris2023honestly} and transfeminine \cite{devito-transfem-tiktok} TikTok creators suggests these groups believe their content is suppressed by TikTok's recommendation algorithms due to their identities. Online communities are also important spaces for people to collectively make sense of representational harms, such as those stemming from algorithmic systems \cite{shen2021everyday}.

    To negotiate between these harms and benefits, research on online communities for marginalized groups often finds that members carefully manage their self-presentation and self-disclosure, a heightened version of the self-disclosure concerns one typically finds in online communities \cite{dimicco2007identity}. For instance, transgender people carefully manage their gender disclosure both within and across social media sites \cite{haimson2015disclosure} and LGBTQ+ people leverage the affordances of online communities to avoid social stigma \cite{devito2018too, blackwell2016lgbt}. At the same time, visibility can benefit marginalized people by helping find allies \cite{blackwell2016lgbt} and build communities \cite{klassen2021more, pinter2021entering, klassen2022blacklives}. As a result, researchers have encouraged online community designers to support selective visibility \cite{pinter2021entering, carrasco2018queer, dym2019coming}. One way to support selective visibility is through anonymity. While anonymity can help members of marginalized groups feel empowered to disclose stigmatized experiences online \cite{leavitt2015throwaway, andalibi2016understanding}, the decreased inhibition of increased anonymity \cite{suler2004online} can also increase harassment \cite{gray2012intersecting}. In sum, marginalized groups must simultaneously negotiate the harms and benefits of visibility in online communities. 

\subsection{Fat Online Communities}

    At the time of our writing, there is relatively little HCI research focusing on the experiences of fat people in online communities, with the notable exception of Jonas et al.'s work on participatory design for fat online information seeking and sharing \cite{jonas2024we}. The authors found online fat information is often poorly organized and does not always account for the intersectional needs of fat people. Beyond HCI, numerous scholars have investigated the ways in which various fat online communities benefit their members. For example, Dickens et al. found that those who participated in the fatopshere perceived it as positively improving both their physical and mental health \cite{dickins2011role}.  Many works have evidenced how online communities are used to challenge dominant narratives of fatness to the benefit of fat people \cite{pause2015rebel, davenport2018size, casado2020m}. For example, researchers have studied the network of fat weblogs that were popular in the early and mid-2000s known as the ``fatosphere.'' Pausé describes the fatosphere as a place where fat people can engage in anti-assimilationist performances which challenge anti-fat expectations of fat people while Casado talks about fat online communities as a place where fat people can ``break the silence''\cite{pause2015rebel, casado2020m}. Davenport et al. find that size acceptance blogs serve to legitimate fat people's lived experiences as a way of knowing in contrast to hegemonic, medical perspectives on fatness and fat people \cite{davenport2018size}. Furthermore, scholarship has shown how members of a private Facebook Groups use it to trade information about doctors and develop communication strategies for their healthcare providers together \cite{kost2023has}. 
    
    At the same time, fat people experience substantial harm online, such as directed hate speech and harassment. For example, \cite{lydecker2016does} found that out of approximately 4500 tweets containing the keyword ``fat,'' the majority of them were negative, often including fat-shaming messages, supporting disordered eating, and that women were more likely to be targeted in these messages than men. Work by Wanniarachchi et al. reinforced the intersection of anti-fatness and misogyny. Through topic modeling and discourse analysis, they show that messages of fat stigma directed at women are more likely to be sexually objectifying and harassing in nature \cite{wanniarachchi2023hate}.
    
    While conducting interviews with body positivity influencers about the microaggressions they receive, Chen et al. found that many of the influencers they interviewed ``accidentally'' became influencers simply by being visible fat people on Instagram, foregrounding the double-edged nature of visibility for fat people \cite{chen2024you}. While scholars like Pausé, Afful, and Webb show how fat people represent themselves online is important in challenging and deconstructing hegemonic notions of fatness, they do not often contend with individual or collective cost of this kind of online participation \cite{pause2015rebel, webb2017fat, afful2015shaping}. As fat scholar Marquisele Mercedes writes, ``achieving fat liberation seems to require that we appear, and that we are seen, in places where this exclusion feels most pointed, like in gyms or restaurants or magazine covers and runways, so on. If the struggle for fat liberation requires that we are seen, what does that mean for those of us who refuse to be looked at?'' \cite{mikey-visibility}.

    This work from humanities and social science researchers has used social media to deepen our understanding of fat people and to document fat oppression. In other words, prior work has used social media as a means through which to study fat people rather than focusing on social computing systems as objects of study. While our work certainly contributes to scholarship invested in understanding fat experiences, our primary contribution as HCI researchers lies in examining the role of technology as a mediator of fat experiences. That is to say, in this work, we turn our attention to the sociotechnical relationships between fat people and the design of online communities with the goal of improving future technology design.

\section{Methods}

\subsection{Positionality Statement}
 
The author team is comprised of multiple individuals who bring different viewpoints, experiences, and skills to this project. A subset of us identify as fat, but we all believe that fat people deserve equitable access to social systems, such as healthcare, employment, and leisure. Moreover, we posit that fat people deserve consentful representation, inclusion, and care in HCI. One of us has experience moderating an online fat community and engaging in fat community organizing. Our positionalities are also shaped by other identities and experiences. We are a mix of US-based and Europe-based researchers with backgrounds in computer science, cultural studies, and law, drawing from a constructivist understanding of our research. Some of us identify as queer and/or disabled. All of the authors are white, which is important to note given prior work on how whiteness often works in opposition to fat liberationist pursuits \cite{pause2021fattening}.

\subsection{Participants and Recruitment}

This work draws on twelve semi-structured interviews in English with fat people, in the summer and fall of 2021. Given this project’s focus on online communities and social media, we recruited via Twitter, Tumblr, Instagram, and Facebook. We asked interested participants to fill out a Qualtrics form which asked for the following information: (1) whether they were at least 18 years old, (2) whether they identify as fat or if there is another word they prefer to use, (3) any other identities or information they wish us to know about them, (4) what language they would like the interview to be in, (5) whether they had any questions for us. Participants had to identify as fat, speak English or German, and be at least 18 years old. Due to our commitment to fat liberation, we intentionally chose to use the word ``fat'' in our call for participation and stated that a subset of the research team are fat themselves. Beyond eligibility criteria, we did not require participants to share any demographic information with us, though various participants did so in the course of this research (cf. Table~\ref{tab:participant_demo}). When referring to individual participants in this paper, we use the pronouns they shared with us.
 
We began each interview asking participants how they prefer to identify specifically. We then used their preferred language, such as ``fat''or ``bigger'', throughout our interview. The majority of our participants (P2, P3, P4, P6, P8, P9, P10) identified themselves as aligned with fat liberation or described themselves as fat activists, but this was not the case for everyone. Some participants disclosed their relative fatness according to the fategories \cite{fategories}. While we do not know the total distribution of our participants across the fategories, there was at least one participant interviewed who fell in each of the small, mid, large and super/infinifat ranges. We would be remiss not to acknowledge the fact that the majority of our participants are white. As such, there are aspects of fat experiences that our work likely under explores. As mentioned above, the whiteness of fat studies scholarship is a well-documented threat to the goals of such scholarship, as anti-fatness is a product of anti-Blackness and colonialism \cite{pause2021fattening, belly-of-the-beast}. Future work should center the experiences of fat Black, Indigenous, and People of Color in online communities.

\begin{table*}[h]
\centering
  \centering
    \ra{1.3}
    \small
    \begin{tabular}{cllll}
    \toprule
    \textbf{PID} & \textbf{Gender} & \textbf{Country} & \textbf{Identities Shared} & \textbf{Online Communities} \\
    \hline
    P0 & Woman & USA & Larger Fat, Asexual/Aromantic, Disabled, Student & Blogs, Facebook, Tumblr, Twitter \\
    % \hline
    P1 & Non-Binary & UK & Overweight, Autistic, Disabled & Facebook, Tumblr, Twitter \\
    % \hline
    P2 & Genderqueer & Canada & Fat, Black, Jewish, Multiply Disabled, Autistic & Instagram, Twitter \\
    % \hline
    P3 & Woman & USA & Fat, Community Activist, Cyclist & Twitter \\
    % \hline
    P4 & Woman & Australia & Fat, Disabled, 30s, Bi, Mom, TikTok Creator  & Facebook, Instagram, TikTok  \\
    % \hline
    P5 & Woman & USA & Fat, Mexican, TikTok Creator & Instagram, TikTok \\
    % \hline
    P6 & Non-Binary & Canada & Fat, Disabled, Student, TikTok Creator & Instagram, TikTok, Twitter \\
    % \hline
    P7 & Woman & Israel & Chubby/Overweight, Student &  Facebook, Twitter \\
    % \hline
    P8 & Woman & USA & Fat, Yoga Instructor & Blogs, Facebook\\
    % \hline
    P9 & Woman & USA & Fat, Student & Instagram, TikTok, Twitter \\
    % \hline
    P10 & Woman &  USA & Fat, Menopausal, Bi, Jewish & Facebook, Twitter \\
    % \hline
    P11 & Woman & USA & Bigger, Medical Student, Runner/Athlete & Facebook, Instagram, TikTok \\
    \bottomrule
    \end{tabular}

\caption{Participant Demographics}    
\label{tab:participant_demo}
\end{table*}

\subsection{Data Collection and Analysis}

Each interview took place over Zoom, and the average interview length was 77 minutes. The Zoom audio recording was collected for transcription, and the video recording was deleted. Participants were compensated for their time with a \$25 USD gift card to a place of their choosing. The first two authors conducted all of the interviews, with the first author leading most interviews and the second author taking notes and suggesting follow-up questions. After asking participants how they relate to the word ``fat,'' the interview protocol then continued to ask participants why they were interested in participating, about the online communities they are involved in and what that involvement looks like, when and how fatness or fat identity comes up in these communities or in other social media use. Given that the interviews were semi-structured, each interview involved asking slightly different questions. However, we typically asked some version of the following questions to each participant:

\begin{itemize}
    \item People use a lot of words to describe fatness. What words do you like to use to describe yourself?
    \item Are there any places online where you talk to other \{preferred language to describe fatness, such as ``fat'' or ``bigger''\} people?
    \item Would you consider \{online community mentioned\} to be a fat accepting space?
    \item If you feel comfortable, can you share any negative experiences you might have had as a fat person online?
    \item What are the best experiences you've had online as a fat person?
    \item If you could change anything about \{particular aspect of online community\} what would you want to change?
\end{itemize}

Each interview was transcribed by the automated software Temi, and then read and corrected by the first two authors as a way to further familiarize ourselves with the data. After individually coding each transcript, the first two authors met to compare codes. In this work, we drew on Boyatzis's approach to thematic analysis \cite{boyatzis1998transforming}. This approach differs from the more commonly used Braun \& Clarke's \cite{braun2006using} in that it is more amenable to a combined top-down and bottom-up approach, utilising both deductive and inductive steps in engaging with the data. Deductively, the first two authors initially coded the transcripts in relation to the research questions presented, categorizing for harms, benefits, and strategies. Then, within these top-level categories, an inductive approach was taken. Quotes were each given a descriptive code, such as ``Harm: ads on Instagram are `clearly not for me'{''} (P2) or ``Benefit: support for dealing with medical discrimination'' (P10). We then engaged in a process of iterative affinity diagramming within these categories to inductively identify patterns among these descriptive codes. For this process, we used the digital white-boarding tool, Miro. For example, numerous benefit sub-codes related to the politics and social construction of knowledge which led us to use feminist consciousness raising as a theoretical lens to make sense of those findings. These themes were then discussed and iterated upon among all authors. As there is relatively little research centering fat people's experiences in HCI, our work is exploratory in nature, hoping to open up possibilities for future lines of inquiry regarding fat people and technology. Hence, it should not be interpreted as making generalizable claims about the experiences of fat people in online communities. Instead, we hope that our findings provide context-rich starting points for future investigations.

\section{Findings}

In the next three sections, we describe how our participants leveraged online communities to engage in consciousness raising around fat identity and mitigate everyday experiences of anti-fatness, such as navigating hostile healthcare systems (Section \ref{sec:benefits}). However, to access these benefits, our participants had to navigate myriad sociotechnical harms, ranging from harassment to discriminatory algorithms (Section \ref{sec:harms}). We then describe the strategies our participants used to negotiate the benefits and harms associated with online communities (Section \ref{sec:strategies}).

\subsection{Benefits}
\label{sec:benefits}

Fat people experience substantial stigma and discrimination. For instance, P8 deplored common public perceptions that fat people are ``poor,'' ``unintelligent,'' and ``lazy.'' P6 described how ``obesity epidemic'' rhetoric scared them as a child, leading them to Google ``is being fat bad?'' and being met with results of ``fear-mongering images of the body.'' P7 believes these messages lead fat people to think they are ``the problem.'' Therefore, fat people must ``manage'' their bodies by pursuing ``prettiness,'' ``thinness,'' and ``fitness'' to be treated with dignity and to access important social systems. Additionally, many participants discussed lacking offline fat friends, family, or community. 

Despite these challenges, our participants were able to meet and discuss shared issues with other fat people through online communities, such as those facilitated on platforms such as Instagram, Twitter, Tumblr, TikTok, Facebook Groups, and on various blogging websites. Through these online communities, our participants were able to develop their understanding of fatness through collective consciousness raising and also mitigate everyday experiences of anti-fatness.

\subsubsection{Understanding Fat Oppression \& Re-Conceptualizing Fatness}
\label{sec:benefits_understand}

Our participants engaged in consciousness raising to make sense of fat oppression and re-conceptualize their own fat identities. They did this by sharing experiences and emotions. Moreover, participants imagined alternative fat futures by sharing and experiencing positive fat representations in online communities.

Participants detailed how their online communities centered discussions of lived experience, following the tradition of feminist consciousness raising sessions. Through their engagement with online fat communities, participants often realized that their experiences of harm and discrimination were shared by others. For example, participants described experiencing substantial medical discrimination (P3, P6, P10). However, through online communities our participants were able to re-conceptualize their individual experiences of discrimination as \textit{systemic} issues by learning from others' stories on social media and engaging in collective sensemaking \cite{mamykina2015collective}. P3, for example, described how social media helped her feel less alone and weather medical illness when it had been wrongfully attributed to her weight:

\begin{quote}
    ``If I didn't have Twitter, I probably would have felt crushed because I didn't get any kind of diagnosis to figure out why I'm so tired. Instead of feeling like I was alone, I kind of felt like, `Nope, this is typical. Every fat person is experiencing this.' So I guess being on Twitter and following other fat people — and especially fat activists — it just gave me a sense of like, `I'm not alone. I'm in solidarity with other fat people.'{''} (P3)
\end{quote}

Additionally, participants engaged in the activity of sharing and validating emotions. Multiple participants described their online fat communities as spaces for ``venting'' (P0, P5, P10) or to `` expel'' (P6) negative emotions after experiencing mistreatment. For example, P7 told us about an instance when she replied to a prominent fat activist on Twitter about her own frustration with her thinner, straight-sized friends complaining about being too fat to her. She shared how affirming it was for the fat activist and other fat people in the replies to share similar experiences and collectively discuss their effects.

Through sharing experiences and emotions, participants were able to re-conceptualize ``the problem of being fat'' not as something intrinsic to them or their bodies, but rather, a system of oppression leveraged against fat people as a social group. For example, P6, an aspiring academic, realized through Instagram and Twitter that fat people can be academics: ``It's not the fat [that's holding them back], it's the other people, you know?'' By assigning blame for the underrepresentation of fat academics to systems of oppression (i.e., ``the other people''), P6 was able to build their self-confidence and imagine their self in a future academic career. P11, a medical student, discussed seeing fat people describe their experiences of medical discrimination, and how witnessing those stories impacted the kind of doctor she wants to be, and the type of care healthcare institutions should provide to fat patients. 

\begin{quote}
 ``When I see that somebody felt dismissed and like they basically didn't get an answer and just got told to lose weight... I couldn't ever imagine saying that to a patient... So [the TikTok videos] definitely impacts me. Don't be that asshole doctor that does that shit. Like on your worst day, still don't be that person.'' (P10) 
\end{quote}

In addition to sharing experiences and feelings, developing alternative political visions of the future is a core tenet of feminist consciousness raising. Participants described how seeing fat people live authentic, joyful lives through visual media in online communities, such as photos and videos, enabled them to envision such futures. This positive representation is highly political as fat people are so often negatively portrayed in media. P4 discussed how the visual nature of TikTok enabled her to imagine living a joyful, happy life as a fat person: ``Seeing joy and seeing rolls and seeing cellulite and just seeing yourself mirrored but not in a negative way.'' Likewise, P0 mentioned how positive fat visual representation in online communities is particularly important for people on the larger end of the fatness spectrum:

\begin{quote}
    ``I think when you finally find a group that either focuses on the large fats or actually manages to balance the different kinds of fat people that exist. It's nice to see pictures of people who look like me, people who are my size and shape doing things, having fun, being happy, wearing cute clothes that they feel good in, doing their hair the way they want and wearing makeup or not as they feel like. That's just something we don't get on a larger scale.'' (P0)
\end{quote}

Nevertheless, participants contended with the limitations of positive visual representation in online communities, which often elevate those with the most amount of bodily privilege (e.g., white, able-bodied, or who have an hour-glass shape). As a highly-visual social movement, most came to question whether body positivity's focus on individual self-acceptance was sufficient to address the needs of fat people as a social group (P0, P3, P4, P6, P8, P9, P10). For instance, P9 described her journey from more body positivity-aligned online communities to those aligned more with fat liberation:   

    \begin{quote}
    
        ``I would probably describe myself as a fat liberationist, and that's different from body positivity because it's looking to completely destroy the systems that allow fatphobia to exist, which also are those same systems allowing racism and ableism and all these other things to exist. Whereas I see body positivity as kind of like working within those systems to like, feel good about your body. That was not the original intent of body positivity. I am aware of that, but I see it now as like watered down to like working within the systems to feel good about yourself.'' (P9)  
    
    \end{quote}    

Additionally, participants represented a diversity of viewpoints regarding the source and operation of fat oppression, often with each participant holding multiple perspectives at the same time. Some were adamant that the medicalization of fatness through the social construction of obesity is the root of anti-fatness (P6), while others expressed the belief that the medicalization of fatness is compatible with weight stigma reduction efforts (P11). Other participants talked about understanding anti-fatness as an extension of anti-Blackness and colonialism (P5), while some talked about the relationship between ableism and anti-fatness (P1). These different foci gesture toward different actions in response to anti-fatness. Much like the participants of 20th century feminist consciousness raising sessions did not arrive at a single shared conception of sexism, our participants did not come to share a singular definition of anti-fatness. Even so, they contended with similar material barriers of a world hostile to fat people. In turn, they used online communities in similar ways to navigate this world.

\subsubsection{Navigating an Anti-Fat World}
\label{sec:benefits_navigate}
 
Through the consciousness raising activities described above, participants were able to conceptualize themselves as part of a marginalized social group that experiences anti-fatness. Participants then realized that they could turn to others online who not only shared similar experiences of anti-fatness but who also interpreted those experiences as oppressive (as opposed to interpreting experiences of anti-fatness as the deserved consequence of being fat). Access to audiences with similar experiences and understandings of anti-fatness enabled our participants to appropriate various online communities to navigate and mitigate experiences of anti-fatness in their own lives. Specifically, our participants used online communities to (1) navigate hostile healthcare systems, (2) seek and share clothing information and (3) supplement inaccessible physical third places.

After realizing that their experiences of medical harm were not individual or one-off experiences, but rather systemic experiences of negligence and discrimination, our participants often turned to online communities to find and discuss strategies for navigating such a hostile healthcare landscape. Participants, such as P6, described how being in community with other fat people gave them access to more accurate, actionable medical information about themselves. As an example, P6 learned from fat online communities that incorrectly sized arm cuffs can misread fat people's blood pressure, which led them to purchase a wrist monitor. Similarly, both P6 and P10 talked about how fat online communities gave them space to discuss their diagnoses with other fat people, learning from collective experience about ``what is normal'' for a fat person with a particular disease. This is important because, for instance, P6 had a broken kneecap go untreated because their doctor misattributed their knee pain to their weight. Moreover, participants often turned to online communities to find weight-neutral or fat-friendly healthcare providers. For example, P3 found their doctor online after learning about the existence of Health at Every Size healthcare providers, and P8 mentioned that people ``looking for recommendations on a doctor that isn't gonna fat shame you'' was one of the most common categories of posts in their fat Facebook Groups. 

Other participants (P4, P6, P10) described learning new strategies to engage with their healthcare providers to receive better care. P10 mentioned learning about ways to engage with medical sexism and ageism when seeking healthcare through participating in a Facebook group for fat people experiencing menopause. Both P4 and P6 described learning that they could ask their healthcare providers to not mention their weight at their appointments, and P6 described learning that they could leave an appointment if a doctor disrespected this boundary. P4 described learning to utilize the following rhetoric when a doctor brought up her weight, which she now recommends to her other fat friends: 

\begin{quote}

    ``I always tell people [to say], `I'll accept if you say that this has to do with weight and if losing weight would help. But how would you treat a thin person?' My fat friends, they're like, `oh my god, I'm really scared to go to the doctor,' And I say, `just say the magic words,' or ask them to put it on your record that they won't treat you unless you lose weight because then if your weight goes downhill, that is on the record. And doctors generally do not want that. So they go out of their way to treat you a lot better if you use those magic words.'' (P4) 

\end{quote}

The knowledge constructed, information shared, and strategies planned within these online fat spaces equipped participants to anticipate and minimize the types of anti-fatness encountered within the healthcare system.

Another prominent experience of anti-fatness voiced among our participants was the issue of finding clothing. While some may take access to clothing for granted, half of our participants (P0, P1, P2, P3, P4, P5, P6) struggled to find clothes that fit on their bodies. As participants pointed out, this is no small inconvenience. P3 mentioned she ``must have black pants'' at work but struggles to find them in her size. The lack of brick and mortar stores which carry plus-size clothing often led participants to use online shopping to find clothes. As P0 puts it: ``if I don't have internet clothes, I don't have clothes, period.''

However, participants still struggled to find clothing when shopping online due to high costs (P1, P6) and few retailers (P2, P3, P6). This led participants to seek out recommendations from fat creators online — commonly on Instagram and in private Facebook Groups. P0 and P4 joined the same Facebook groups for fat people with a particular aesthetic where members are encouraged to share where they shop. P4 discussed joining another Facebook group for fat sewists to learn how to make her own clothing. P6 reached out to a prominent ``large fat'' Instagram influencer to ask where they bought a sundress and was overjoyed when the creator responded with a link.

Finally, online communities were also noted as particularly important for our participants because offline third places are often not physically accessible or socially safe. For instance, the design of ostensibly mundane objects like chairs can render physical meeting spaces inaccessible. Both P2 and P8 discussed the problem with chairs not fitting them at restaurants, nail salons, and on planes, while P0 described not fitting into chairs at her university. In addition to physical accessibility, public spaces can also be socially inaccessible. P8 recalled the challenges she faced trying to meet in-person with members from her local fat online community due to discrimination:

\begin{quote}

    ``We [members of this fat online community] would go out and do something social. But that kind of dissolved when the manager of the club really didn't like the idea of having a bunch of fat women come to the club. He wanted thin women coming to the club. Even if we paid ahead of time to have a separate room, we weren't able to get another place to allow us in.'' (P8)
    
\end{quote}

In contrast, online communities are less bound to the constraints of physical meeting spaces or in-person social gatekeepers. Our participants described fat online communities as safe and accepting places to connect with others, form friendships, learn together, and network professionally. What are often considered the benefits of ``third places,'' \cite{oldenburg1999great} fat people access through online communities. P8, a yoga instructor who focuses on making yoga accessible to fat people, described participating in an online Facebook group for fat/curvy yoga instructors as a place to learn and connect with similar professionals without the hostility she had encountered from thin yoga professionals in offline professional networking events. Similarly, P9, a fat student studying nutrition, described starting her own online fat study group with a friend over Zoom and Instagram, noting that there were no offline spaces for them to do so. Each of these participants, as well as others, also emphasized that they had made many friendships through their participation in these online communities. While participants would later complicate the notion of these online spaces being fat-accepting (as we explore in the section on harms below), many participants agreed that online communities were more inclusive and welcoming than offline spaces. As P1 said, ``I find online is probably more accepting than real life.'' 

\subsection{Harms}
\label{sec:harms}

\begin{quote}
    \textit{Content Warning: The following section includes descriptions of harassment our participants experienced, including examples of hateful language directed at them. We encourage readers to proceed with care.}    
\end{quote}

While participants undoubtedly benefit from these online fat communities, participants expressed a variety of harms they experience in accessing these spaces and outlined the many ways in which social media platforms uphold anti-fatness. As P3 said, ``I can exist a little bit online and escape some of the discrimination of being fat, but not for that long. It's always out there in some form.'' Below, we discuss participants' experiences of trolling, harassment, and silencing online, failures of content moderation systems to meet the needs of fat people, and various representational harms which reinforce the devaluation of fat people.

Each of the harms discussed below are sociotechnical in nature. Throughout our interviews, we often asked participants if they thought a particular platform they were discussing was ``fat-accepting.'' When discussing Twitter, P2 responded with, ``I don't think that the forum is the problem. I think the people are the problem.'' This was a common response among participants when asked. At the same time, participants often discussed the design of social media platforms as a source of suffering, which we highlight below when relevant. We also outline the ways in which the harms discussed below are often mutually reinforcing, connected by similar actors, technical systems, and beliefs about fat people. 

\subsubsection{Harassment and Anti-Fat Content Moderation}
\label{sec:harm_content}

    Perhaps the most common negative experiences participants shared with us were of being trolled and harassed online (P0, P1, P2, P3, P4, P6, P9, P10, P11). P4 described the breadth of harassment she receives in response to her TikTok videos: 

    \begin{quote}
        “So it ranges from like, ‘oh my God, she's breathing so heavily, what a fat troll, what a fat whale, disgusting’ kind of comments to the people who pretend like they care about you and your health. You know, the people who are like, ‘I just don't think it's healthy. I just don't think it's good to promote obesity.’ I treat them pretty much the same, the people who think that they are trying to help. They don't care. So I'm not going to engage with it. It’s actually… sea lioning.” (P4)
    \end{quote}
    
    ``Sea lioning'' and ``concern trolling'' were common forms of harassment participants described \cite{jhaver2018online}, where trolls would leverage bad faith questions or ``concerns'' about a participant's health. However, other experiences of harassment were not as thinly veiled and often targeted multiple aspects of a participant's identity. P2, a trans, non-binary Black person described how the harassment they receive is a product of anti-fatness and racism: ``If I get trolled, I get trolled hard. I don't really get trolled with like, `please stay healthy.’ I get trolled with like, ‘go to die.’”  
    
    In addition to experiencing direct trolling, participants described the harm of seeing other fat people trolled or harassed. P3 mentioned the harm of incidentally encountering ``fat hatred'' when ostensibly progressive friends make fun of fat conservative politicians online. Similarly, P2 mentioned: ``there's an aphorism about how the famous person that you're tweeting [at], won't ever hear you... but all your fat friends will.'' 
    
    Seeing other fat people harassed online led to a silencing effect for many participants, especially with regards to photo sharing. In particular, participants were afraid of harassment that might occur when photos of themselves are appropriated by anti-fat actors. P1, a non-binary participant, shared a story about posting a haircut photo on Twitter, and being quote-tweeted with harassing remarks. Furthermore, P3 discussed her fear about where her photos might end up after posting: ``I don't want my photos to be retweeted or shared in a negative light, and especially not to promote eating disorders. There's definitely those corners of Twitter that are not fat friendly.''

    Exacerbating these harms, participants mentioned that trolling and harassment toward fat people is often under-moderated. Both P3 and P6 mentioned that online communities do not consider fat hate to be as serious as other forms of hate. When discussing harassment, P3 said, “I would like Twitter to treat fat hate as the way they treat other hate online.'' Likewise, P6 tried to report fat hate on Twitter as ``hate or discrimination towards a protected group'' but shared they ``quickly found out that we [fat people] are not a protected group.''  This may be why fat hate is omnipresent in online communities: ``When I come across a fat person posting, there's usually fat hate underneath it'' (P9).

    In addition to what participants perceived as \textit{false negatives} in content moderation (harassing content not being removed), many had the experience of \textit{false positives} in that their own content was over-moderated and wrongfully taken down. Both P4 and P5 described their experiences creating content for TikTok, and having their videos flagged as violating TikTok’s nudity policy. P4 worried that “if they [fat content creators] jiggle a bit too much, then they risk having their account being Thanos-ed [gone at the snap of a finger].” In the context of TikTok, where views are correlated with opportunities for monetization, P4 raised concerns about the potential financial harms associated with this kind of misclassification. P5 described a similar personal experience: 

    \begin{quote}
        “A video would be pushed out and pulled down immediately for nudity. And I would just be wearing a top that would show skin. And [sarcasm] when you’re fat, skin is nudity, who would've thunk… Fat creators [have to be] hypervigilant about how much skin they share. So while you see 10 people in tiny little bikinis, doing little dances, shaking their little booties, a fat person does the same thing and they would be Thanos-ed. Or at the very least the video will go under review, sometimes for several days. Sometimes it will come back, sometimes it won’t. And that happens to almost every fat creator that I know.''
    \end{quote}

    In sum, our participants described experiences of substantial harassment, which were, at best, ignored by content moderation systems and, at worst, mistakenly led to them becoming targets of these systems.

\subsubsection{Representational Harms}
\label{sec:harm_representation}

In addition to harms resulting from harassment and content moderation, we found that anti-fatness is embedded in other aspects of the design of online communities. Participants highlighted a number of representational harms. In the context of our inquiry, we define ``fat representational harms'' as those occurring when the design of a technology devalues fat people, perpetuates harmful stereotypes about fat people, or erases fat people from media artifacts entirely.

%  Filters
P2 and P4 discussed the representational harms associated with particular ``filters” on platforms including Snapchat, TikTok, or Instagram. A ``filter'' is an augmented reality video effect which can alter the appearance of oneself or one's environment, among other effects. P4 discussed the pain she felt watching thin people use a ``fat face” filter on TikTok and, upon removing it, make statements like ``oh, thank god I don't look like that.'' P2 imagined a potential future where filtering technologies do not associate being thin with being better:  “In some of the `bettering you' filtering technologies, maybe don't make me skinnier in all of the filters, maybe some of the filters. It's a variant of humanity, but maybe everything doesn't have to make me seem thinner for the world.'' However, for now, participants felt like TikTok filters upheld the idea that being fat was bad (P2, P4). 

% Gifs
P7 mentioned another representational harm: a GIF-search tool embedded in a Discord server that reinforces negative stereotypes about fat people. She describes how GIF-representations of fat people appear in searches for comedic terms but not for other terms, such as ``kiss'' or ``beach.'' In doing so, the design of this GIF-search tool upholds the well-explored notion that fat people are consigned to particular roles and not others, such as the funny sidekick or comedic relief, as opposed to a desirable romantic partner. P7 went on to discuss a general lack of fat representation in social media discourse, which negatively impacted her self-image:
    
    \begin{quote}
        ``Maybe [it’s] just because I don't follow a lot of fat creators, but most of the content that I see, be it photos or art, or GIFs from TV series, everyone [is] just kind of by default thin, or slim. And you just kind of take that as that's just how the world is. Like, everyone looks like that. Even start to think that you look like that, and then when you go offline and you look at yourself in the mirror and you're like, oh no, I don't look like that.'' (P7)
    \end{quote}

% Programatic advertsiing
P2 and P3 provided examples of another kind of representational harm: fat people being excluded from online advertising. Both discussed their experiences with programmatic advertising, and the way in which the ads they receive make them feel like platforms and advertisers are imagining a world without fat people. P3 mentioned the disappointment of clicking on clothing related banner ads only to find no plus-size clothing options. P2 further elaborated on the harm of receiving Instagram ads that are ``clearly not for [them]:'' 

    \begin{quote}
        ``I get ads for clothes I couldn't possibly wear. I get ads for shoes I couldn't possibly wear. I get ads for things I might be interested in, but all of the models are white or the ad copy is trans[phobic] or fatmisic in a way that I just don't want to engage with.'' (P2)
    \end{quote}

Across these examples of representational harms, we see participants noticing how various features of online communities are designed in ways that stereotype and fail to consider fat people.

\subsection{Strategies}
\label{sec:strategies}

Throughout their interviews, participants identified a number of strategies they use to maximize the benefits and minimize the harms they encounter online. Here, we present two primary categories of strategies: staying silent and opting out, and speaking up and making space. Participants tend to adopt staying silent or opting out of specific platforms as a means to protect themselves from the kinds of harms we outline above. At the same time, participants often risk these harms by speaking out on social media in an effort to further raise fat consciousness. The combination of these two strategies attends to the sociotechnical nature of the harms and benefits participants receive, as they are mindful of both technical systems as well as human actors. It is worth mentioning, too, that the labor of deploying these strategies may constitute subsequent harm. 

\subsubsection{Content Moderation, Counter-Discourse \& Care}
\label{sec:strategy_counter}

To combat trolling, harassment, and anti-fat discourses, participants turned to safety tools provided by platforms hosting online communities. For example, P0 and P3 discussed blocking trolls, while P3 and P6 mentioned reporting harassing accounts to various platforms. However, these tools have limited efficacy. P10 worried, ``even if you block somebody, if they know your first name and your last name and other information about you, they can come find you.''  P3 and P6 described the harm of anti-fat hate not fitting into the schema of existing content moderation reporting systems. As they were unable to report anti-fat content as discrimination against a protected group, P3 and P6 reported anti-fat content on Twitter and Instagram as misinformation (P6) and bullying (P3). Moreover, the potential of being harmed by content moderation systems led P4 and P5 to engage in archival work for fear of their accounts or individual videos being banned by TikTok. For instance, P5 mentioned screen recording the TikTok videos she makes because posting videos feels like ``a gamble'' on whether the videos will be taken down. P2 further emphasized how these individualized safety tools may fail to address the collective impact of anti-fat comments on on other fat people: 

  \begin{quote}
        ``I am not an avid user of the block. I tend to just ignore people and let them keep screaming into the void. But if I think that what they're saying is particularly harmful, and I know that my audience is going to see it, I'm going to respond in a way that offers support to other people who are going to be hurt worse by what that person is saying.'' (P2)
    \end{quote}

Given the limitations of these safety tools, participants often engaged in strategies of speaking up and making space for other fat people online. Participants turned to various counter-discursive strategies — such as humor, sharing personal stories, and debunking  — to combat both directed anti-fat hate and more general anti-fat discourses. In so doing, participants sought to care for themselves as well as other fat people. For example, P4 satirizes or makes fun of her trolls on TikTok as a way to diminish their power: ``I will make a video reply with, you know, a meme. Just to show that I'm unbothered.'' Some participants mentioned debunking anti-fat claims as part of their typical online activity (P2, P3, P6, P10). Because trolls often use academic and medical sources to justify anti-fat harassment, P2 and P6 described carefully researching, curating, and organizing academic and journalistic sources to have ``on hand for talking to people and trying to educate people'' (P6). In response to trolls telling P3 she is ``going to get COVID and die'' because she is fat, P3 mentioned being ``so thankful for that Wired article that came out last year because I can just paste that and be like, `Nope, that's been debunked by Wired months and months ago.'{''}  Moreover, P10 voiced that she hoped to ``help support somebody, even if they're a stranger'' in re-conceptualizing the way they think about fatness by ``dispelling some of the BS around fatness, fat bodies, and fat people.'' P2 and P6 described sharing instances of offline anti-fatness online as a way to make anti-fat discrimination more legible to others. P2 shared the following anecdote with us: 

\begin{quote}
    ``I bring a lot of real life situations back to Twitter-verse. If something happens to me in a space, for example, when I got my first dose of [the COVID-19] vaccine, they brought me to the room where you're supposed to wait for 15 minutes before you leave. I got like the only chair for fat people. I looked around and every chair had arms and every chair was a static situation that you couldn't really swap out or move. So if you are the fat person, you have to sit in this one spot that is maybe not where you want to be. And so that was something that I brought back to Twitter and was like, `Hey, so here's my experience. Here's how it could be nice to consider fat people in the world, in everything, not just when you're doing fat things.' I think that I'm not alone in bringing the real world back [to share on social media].'' (P2)
\end{quote}

Of course, these counter-discursive strategies are invariably entangled with the consciousness raising activities we described above.

Finally, participants also engaged in strategies of care with one another after experiencing trolling and harassment, usually through one-on-one direct messaging. P6 described messaging another fat person on Instagram after seeing hateful, anti-fat comments directed at her. P5 spoke to the importance of direct messages as sites of care, saying: 

\begin{quote}
    ``A lot of it takes place in like DMs. Like you have no idea how like open people are... people will just tell you their entire life story over a DM. And you're like, `I'm listening. You want me to give you advice? You want me to give you a solution? You want me to put you in contact with somebody else? Let me know what's up. I can help.''' (P5) 
\end{quote}

Many of the strategies participants shared with us around dealing with harassment prioritized the impact those strategies would have on other fat people sharing the same digital space. However, participants also showed care and consideration for their own safety, which we discuss in the next section. 

\subsubsection{Identity Management, Leaving \& Building Community}
\label{sec:strategy_visibility}

Participants carefully managed the disclosure of their fatness in online communities for fear of trolls. For instance, P1 and P3 mentioned rarely posting photos of themselves on Twitter for fear of trolls, while P2 shared that they generally only post head shots of themself online. However, on short-form video platforms like TikTok, creators may not have the luxury of choosing when to disclose one's body. Two of the fat TikTok creators we interviewed (P4, P5) mentioned taking great care about how they displayed their body when posting for fear of their videos being misclassified as violating TikTok's community guidelines around nudity. While P1, P2 and P3 carefully manage their self-disclosure due to harm from an imagined audience of other social media users, P4 and P5 negotiate their self-disclosure due to both the possibility of harm from an imagined, opaque content moderation systems. 

As the ultimate form of identity management, some participants chose to opt-out of entire online communities or platforms. For instance, P9 mentioned that she would never post on TikTok due to the limited ability to manage one's audience: ``I personally wouldn't post on TikTok because I would just feel like it could pop up on anyone's [For You Page] and like easily become like this troll-fest.'' P2 opted out of Reddit because: ``Reddit is hell. Reddit is where all the hate is.'' Participants discussed leaving or considering leaving online communities. P5, whose TikTok videos were wrongly flagged for containing nudity, mentioned they were considering leaving the platform and transitioning to YouTube. 

Leaving and opting out were not only strategies for dealing with harassment and unfair content moderation systems, but also for dealing with systems of oppression reproduced within online fat spaces. P5, who is an Indigenous Mexican-American woman, opted out of online fat discourses as ``a lot of conversations are being led by white women.'' P1 discussed leaving online body positivity communities, saying they felt ``left out'' by the focus on self-acceptance and beauty which minimizes how fatness impacts their life, such as being denied gender-affirming care due to their weight. Likewise, P4 came to critique that ``white bodies,'' ``small fat bodies,'' and ``hourglass bodies'' were at the forefront of the body positivity movement. P0 mentioned how larger fat people, like herself, often get ``steamrolled by the smaller fats'' in online communities, ultimately driving her to leave a private, fat-centered Facebook group:  

    \begin{quote}
    
        “And to be super blunt, I got sick of the fat, the small ones, medium fats whining, and just kind of refusing to acknowledge that they have very legitimate problems and lack of privilege, but not to the extent that people like me do who are larger fats... and they would get offended every single time the larger fats would say, ‘Hey, maybe do you think we could like amplify larger fat voices occasionally?’ And I felt like the moderators and weren't actually doing anything about it. So I just finally said, `this is just increasing my blood pressure, I'm leaving.' And I did.” (P0)
        
    \end{quote}

Finally, participants described making their own fat accepting spaces across a variety of mediums. For example, P5 discussed building community in the comments section on her TikTok videos, and as mentioned above, P9 and some of her friends created an Instagram page and Zoom study group for healthcare students interested in fat liberation.

\section{Discussion}

In this section, we discuss how HCI researchers can reconsider how they know and design for fat people. We then demonstrate how this can be enacted by identifying particular implications for the design of online communities that are raised by our findings. Much like the HCI community is increasingly expanding the ``humans'' we consider when studying human-computer interaction, we contend that this effort should also extend to fat people. As we will explain, designing for/with fat people requires contending with the material realities of fat embodiment, the social realities of fat identity, and the materio-social nature of fat oppression. In doing so, our work is entangled with other social justice efforts in HCI advocating for rethinking normative assumptions about users and their bodies \cite{keyes2020reimagining, morris2023don, freeman2022acting, baumer2017post}.

\subsection{Knowing and Designing for Fat People}

        In reflecting on our research process, we can see that the lens we chose to study fat people — as a marginalized social group with shared experiences — directly impacted the nature of our findings. For instance, one cannot identify representational harms toward fat people without first conceptualizing fat people as a social group that can be negatively represented (Section \ref{sec:harm_representation}). Moreover, one cannot understand how our participants used online communities to navigate offline anti-fatness (e.g., searching for clothing and less discriminatory doctors) without recognizing the systemic nature of anti-fatness (Section \ref{sec:benefits_navigate}). Solely studying fat people through the lens of weight-loss risks overlooking these findings. \textbf{It follows that HCI researchers can neither fully know fat people nor know marginalization without contending with fat people as a marginalized social group.} While we adopted this perspective in our work, we still found substantial heterogeneity in how our participants self-identified (e.g., as ``fat,'' ``overweight,'' or ``bigger''), made sense of their experiences (e.g., fat liberation, weight stigma, or body positivity), and experienced fatness in relation to other aspects of their identities (e.g., race, gender, or disability). These factors mediated our participants' experiences as fat people online. For instance, P5 — an Indigenous Mexican-American woman — was turned off from participating in fat spaces led by white women (Section \ref{sec:strategy_counter}). P1 was displeased with the body positivity community's focus on self-acceptance and reproduction of hegemonic beauty standards, which does not speak to their experiences as a fat, disabled, trans person. Our findings highlight the need to incorporate fatness within intersectional analyses of identity and marginalization in HCI.

        In order to understand fat people as a marginalized group, we encourage HCI researchers to regard fat people as the utmost experts on their own lived experiences \cite{cooper2020elevating, pause2020ray, manokaran2021nothing}. For example, our participants described experiencing inadequate medical treatment, such as having injuries not believed by doctors who misattributed pain to fatness (Section \ref{sec:benefits_navigate}). It follows that those technologists who take medical expertise for granted risk reproducing and exacerbating extant epistemic injustice \cite{fricker2007epistemic}. Our participants also shared stories of how both fat oppression and fat liberation can be facilitated through journalistic and scholarly research. Multiple participants described trolls using academic research to justify their harassment (Section \ref{sec:harm_content}) and, in turn, described maintaining archives of academic and journalistic research to support their counter-discourse against trolls and to help raise the consciousness of others (Section \ref{sec:strategy_counter}). In other words, academic research can be either part of the problem or part of the solution to anti-fatness. Much like our participants raised their consciousnesses around fatness through online communities, we suggest that we as HCI researchers also need our consciousnesses raised. Centering and consentfully engaging with the expertise held by fat people about their own lives is one avenue toward consciousness raising within HCI. Here, we see parallels between our findings and related efforts in HCI advocating for the agency of disabled people \cite{bennett2019promise}, those experiencing homelessness \cite{kuo2023understanding}, and autistic people \cite{spiel2019agency, guberman2023actuallyautistic}.

        In future research, we call on HCI researchers to interrogate the ways in which anti-fatness is embedded — both explicitly and implicitly — as a value in the design of technology. As can be seen in recent research on other marginalized groups in HCI — such as LGBTQ+ people \cite{taylor2024cruising} and People of Color \cite{benjamin2019race} — it is likely that designers will perpetuate anti-fatness if they do not explicitly consider fat people in design. The harm of ignoring anti-fatness in design can be seen in our participants' experiences navigating trolls. Our participants recognized that the anti-fat harassment they experienced was not considered hate speech because fat people are not considered a protected class by online community designers. This inability to report anti-fat harassment as hate speech reflects how anti-fatness has often been relegated to a residual category of harassment or discrimination. Here, one can see parallels between our work and some trans HCI research, such as the exclusion of non-binary people from everyday technological classification infrastructure \cite{spiel2021they} and erasure of transgender people from automated gender recognition systems \cite{hamidi2018gender}. The designers of algorithmic systems likely do not intend to discriminate against Black, trans, or fat people. Rather, designers will reify extant racism, transphobia, and anti-fatness unless they undertake deliberate efforts to resist these systems of oppression. Moreover, fatness intersects with other identities such that it is impossible for designers to fully consider, for instance, communities of color or LGBTQ+ people in design without also considering the impacts of anti-fatness. 

        At the same time, we encourage researchers to approach the study of fat people as a marginalized group with care. There is ongoing debate among those researching marginalized groups over the extent to which HCI researchers should center harms, deficits, or damage in their work relative to joy or everyday experiences \cite{taylor2024carefully, cunningham2023grounds}. Certainly, the anti-fatness associated with online harassment and representational harms are important to address and warrant further research attention in HCI (Section \ref{sec:harms}). At the same time, studying fat people requires more than studying and trying to mitigate anti-fatness. We also encourage researchers to focus on helping fat people to — borrowing from To et al.'s work on the topic — ``flourish in the everyday'' \cite{to2023flourishing}. That is to say, the experience P0 described of seeing fat people positively represented on social media deserves attention alongside studies of anti-fatness. Future research should consider both how to reduce harm and increase joy while continuing to explore how the harms and benefits of online communities may be in tension with one another.

    \subsection{Designing Better Online Communities for Fat People}
    
    In order to confront anti-fatness in design, \textbf{we must contend with both material and social facets of fat experience}. At times, this may involve centering the materiality of fat embodiment in design while, in other circumstances, contending with the myriad ways in which fatness is socially constructed. Here, we see similarities between Harrison \& Dourish's famous distinction between space (the material configurations of a location) and place (the social and cultural context of a space) in HCI \cite{harrison1996re}. For example, a ``space'' may be physically inaccessible to fat people due to its material configuration (e.g., size of chairs), while a ``place'' may be socially inaccessible due to the behavior of other people (e.g., fat harassment). In other words, our findings emphasize the importance of contending with the materiality of bodies, how individuals make sense of their bodies, and how designers make sense of peoples' bodies \cite{fox2020multiples}. Below, we elaborate on what it looks like to consider the material and social aspects of fat experience in the design of online communities. In this subsection, we discuss implications for research and design related to fat people and online communities.
    
    \subsubsection{Improve Content Moderation}

    One site where designers must consider the material aspects of fat experience is in the context of visual content moderation algorithms. Our participants described the harms associated with being misclassified by TikTok's NSFW algorithmic content moderation system. Auditing this algorithmic system for anti-fat bias would require prioritizing the material experiences of anti-fatness. This is because — although identification and appearance may be correlated — this algorithmic discrimination primarily results from how one looks rather than how one identifies. Although work has extensively examined racial and gender biases in computer vision models \cite{buolamwini2018gender, scheuerman2019computers}, comparatively little work has examined anti-fatness.

    Meanwhile, attending to the social construction of fatness is also crucial for addressing the substantial harassment our participants described experiencing in online communities. Our participants noted how anti-fat harassment fell outside of the classification schemes embedded in social media content moderation systems. Our participants were unable to report anti-fat content as hate speech because fat people are not considered a protected class, even though fatness is highly racialized and gendered \cite{belly-of-the-beast}. While previous investigative reporting such as \cite{invisible_censorship, instagram_fat_discrimination} uncover both implicit and explicit anti-fat content moderation policies on platforms such as TikTok and Instagram, here we point out another way in which anti-fat content moderation is facilitated: the maintenance of fat people's status as a residual category. To address this harm, those seeking to design a content moderation policy or a user interface to report fat hate must contend with the myriad, contested ways in which fatness is socially understood by those who may file and moderate reports. How can one protect a group whose members may not see themselves as belonging to a group? Future research might also explore whether fat people are impacted by linguistic discrimination, as the word ``fat'' can either be used as a neutral descriptor or an epithet depending on how it is used. For example, toxicity detection algorithms \cite{thiago2021fighting} and moderation block lists \cite{dodge2021documenting} often discriminate against LGBTQ+ people due to in-community language and reclaimed slurs. While explorations of visual algorithmic discrimination might focus more on the material aspects of fatness, exploration of linguistic discrimination or content moderation classification design focuses more on the social construction of fatness.
            
    To remedy the resulting harm of anti-fat content being under-moderated due to falling outside classification systems, we encourage designers to \textbf{add anti-fat harassment to hate speech content moderation schemas}. At the same time, we caution against overzealous moderation — such as the NSFW TikTok false positives in Section \ref{sec:harms} — that may silence those a policy is intended to protect \cite{haimson2021disproportionate}. Furthermore, the privilege held by different fat people (Section \ref{sec:strategy_visibility}) demonstrates the need for designers of algorithmic systems to consider intra-community power dynamics \cite{walker2020more} and intersectionality when moderating anti-fat content \cite{gilbert2023towards}. Taking down content may also not be the best practice in all situations \cite{xiao2023addressing}. For instance, Salehi et al. questioned the efficacy of punitive content moderation to address the harms Muslim Americans experience in online communities, instead advocating for restorative justice approaches \cite{salehi2023sustained}. Future research should explore fat people's content moderation preferences in greater detail. 

    \subsubsection{Support Selective Fat Visibility}
        
    Our findings point to both substantial harms and benefits associated with fat people disclosing their fatness online. Some participants worried about sharing images or videos of their body online for fear of harassment (Section \ref{sec:harm_content}). At the same time, participants also emphasized the pleasure of sharing photos of oneself online and the benefits of seeing other fat people visible online (Section \ref{sec:benefits_understand}). Due to the systemic nature of anti-fatness, visibility makes fat people vulnerable online. However, as Barta et al. remind us \cite{barta2023toward}, vulnerability can be both positive and negative, opening oneself up to both pleasure and pain. To help fat people negotiate the harms and benefits associated with visibility, \textbf{we encourage designers of online communities to better support selective visibility}. For instance, P9 decided not to post on TikTok for fear of her content being algorithmically shown to anti-fat harassers, which could be remedied by improved audience management tools. Moreover, supporting fat selective visibility likely involves allowing — but not obliging — users to share their ``real body'' in online communities. For instance, virtual reality communities should allow — but not require — fat people to create fat avatars to represent themselves. 
    
    Our suggestion here parallels the tension noted by Schulenberg et al. in relation to women's experiences in Social VR \cite{schulenberg2023creepy}: disguising one's marginalized identity may help prevent individual harassment, while at the same time leading to collective harm by reducing online representation. One potential way to address this tension is supporting less binary forms of disclosure, paralleling how LGBTQ+ parents use Facebook profile badges to identify allies within their social network \cite{blackwell2016lgbt}. Another strategy may be creating closed online communities in which it is safer to be visible, like the closed Facebook groups described by P0. These communities could be maintained through photo verification — such as the practice of verifying Black users within the r/BlackPeopleTwitter subreddit \cite{smith2024governance} — but this may also risk essentializing fatness. In advocating for selective visibility, our recommendations echo prior work on designing online communities to support other marginalized groups \cite{pinter2021entering, carrasco2018queer}. While our work suggests that fatness should be included within these marginalized disclosure considerations, an open question remains as to how designers and researchers can support the benefits of visibility while also mitigating the harms of visibility.

    \subsubsection{Design for Fat Community Building}

    Based on our findings, we suggest that designing online communities to support consciousness raising — particularly for members of marginalized groups — involves \textbf{supporting boundary management}. As we mentioned in Section \ref{sec:benefits}, private Facebook Groups gave some of our participants the freedom to discuss fat issues with less fear of encountering anti-fat hate. Likewise, P9 worried about joining TikTok for fear of her content being shown to trolls. It follows that one can better support fat consciousness raising by empowering fat people to construct their own spaces. Moreover — in light of the harms our participants mentioned related to TikTok, Instagram, and Twitter's global moderation practices — fat online communities may benefit from more local moderation. Our recommendations here echo similar concerns mentioned in prior work on online communities for marginalized groups, such as concerns about outsiders on Black Twitter \cite{klassen2021more} and the need for local moderation practices in r/bisexual \cite{taylor2024mitigating}. However, boundary management lies in tension with the desire for fat people to be able to discover fat online communities. Future work should explore how to help fat people find existing online communities while also keeping out trolls or interlopers.

    HCI researchers should explore how they might support fat people \textbf{build online communities outside dominant social media platforms}. This is particularly prescient as Twitter has become more hostile toward marginalized groups since our interviews took place \cite{marginalized_twitter_elon} and US content creators are bracing for a potential TikTok ban \cite{tiktok_ban}. This may involve designing online communities that embed fat liberationist values into their design from the ground up, paralleling how fanfiction writings embedded feminist principles in the design of AO3 \cite{fiesler2016archive}.\textbf{ Incorporating fat liberationist values into the design of online communities} would require researchers and designers to consider the fattest among us first, as D'Ignazio et al. point out that ``intenionally designing for equity'' is a quality which supports feminist consciousness raising \cite{d2020personal}. Likewise, the design of fat online communities should prioritize storytelling to support consciousness raising. This recommendation follows prior work on the importance of storytelling for building online social movements \cite{dimond2013hollaback, jackson2020hashtagactivism}. Building on HCI research on the design of trans technologies \cite{haimson2020designing} and supporting those experiencing street harassment \cite{dimond2013hollaback}, fat online communities may also benefit from the ability to make counter-maps to help participants find doctors (Section \ref{sec:benefits_navigate}). These online communities may also help remediate harms associated with other platforms (Section \ref{sec:harms}), such as the emotional support provided by HeartMob \cite{blackwell2017classification} or the material support provided by blocklists remediate harms on other platforms \cite{jhaver2018online}. HCI researchers should also explore how they might support fat community building and discourse beyond online communities. Much like DiSalvo et al. highlight the potential for design to facilitate the formation of publics \cite{disalvo2014making}, so too should HCI researchers consider how to facilitate the formation of fat counterpublics.
    
    Finally, we caution HCI researchers not to forget that fat people are and have long been creating their own online and offline infrastructures \cite{soden2021time}. We call on HCI researchers to \textbf{recognize and learn from existing fat infrastructuring practices} \cite{semaan2019routine}. Fat people are already engaged in building their own infrastructure, such as the catalogs of research about fatness mentioned by P2 and P6 (Section \ref{sec:strategy_counter}), or artifacts like a Google Sheets list which houses an extensive list of recent plus-size clothing retailers \footnote{A document we have seen but chose not to link for privacy reasons.} or the Fat Liberation Archive which hosts a digitized repository of historical documents related to fat activism. \footnote{The Fat Liberation Archive https://fatlibarchive.org/} In investigating fat online infrastructure, researchers must also work to surface the invisible human labor undergirding technological infrastructure \cite{britton2019mothers}. Again, doing this work will require researchers to engage with fat people as experts on fat experiences.

\section{Conclusion}

The 1983 anthology \textit{Shadow on a Tightrope: Writings by Women on Fat Oppression} includes a series of essays reflecting on the ``first'' ten years of fat activism organizing in the United States. In this collection, early fat activist and Fat Underground founder Vivian Mayer wrote about her experience as a fat organizer in California - the way small groups of women came together to discuss their experiences, oppression, and to find joy in each other's company. In her writing, she reflected on the isolation groups of fat people faced in the 1970s through both epistemic and geographic separation \cite{schoenfielder1983shadow}. But she also wrote about hope for the future of fat organizing, saying: 

\begin{quote}
    There are signs that the isolation is ending, and that fat women’s liberation will soon be recognized as a distinct yet fundamental feminist voice. As this happens, I hope that fat women who struggled to build this movement in the past will come forward to share their experiences and thoughts with new activists… I pray for an outpouring of women’s thought, speech, and action to come soon, in our days.” 
    
    --Vivian Mayer, Storrs, Connecticut, April 1983 
\end{quote}

In our exploratory study, we found that online communities are helping to reduce the epistemic and geographic distance between fat people and provide space for their thoughts, speech, and actions. Through online consciousness raising, participants were able to re-conceptualize harmful, hegemonic notions of fatness. Armed with this understanding, our participants were able to raise the consciousness of others and seek and provide help navigating an anti-fat world. However, our participants also reported that engaging in online communities comes at cost, experiencing harms from both other users and directly from the design of online communities themselves. In turn, participants deployed a number of strategies, ranging from lampooning trolls to leaving platforms entirely.

Based on these findings, we encourage researchers and designers to consider both how they engage with fat people more broadly as well as the specific ways to attend to fat people in the design of online communities. These include improving content moderation systems to better protect fat people, supporting selective visibility, and engaging with existing fat communities and their ongoing infrastructuring practices. More broadly, we hope that our work provides openings to explore more equitable fat futures in HCI.

\begin{acks}

We would like to extend our deepest thanks to our participants for sharing their time and stories with us. We would also like to thank the following people for their early feedback on this project: Kendra Albert, Stevie Chancellor, Jes Feuston, Jessica Hammer, Lynn Dombrowski, Munmun De Choudhury and the Social Dynamics and Wellbeing Lab. Thank you, too, to the Al-Adala Lab and Internet Rules Lab at CU Boulder for their feedback at all stages of this work. The first author would also like to thank Rachel Fox, Marquisele Mercedes, and Monica Kriete for their support and fat solidarity. The second author would like to thank Haiyi Zhu and Sarah Fox along with the members the Social AI and Tech Solidarity Labs for their support. Finally, thank you to our anonymous reviewers, whose thoughtful feedback greatly improved this work. This work was supported by the National Science Foundation (\#1704303).

\end{acks}

\bibliographystyle{ACM-Reference-Format}
\bibliography{refs}

\end{document}